\documentclass[prd,nofootinbib,showpacs,preprint]{revtex4}
\usepackage{amsmath}
\usepackage{graphicx}
\usepackage{epsfig}
\usepackage{dcolumn}
\usepackage{bm}
\usepackage{amssymb}
\usepackage[usenames,dvipsnames]{color}
\usepackage{slashed}
\usepackage[dvipdfm,colorlinks,citecolor=blue]{hyperref}

\begin{document}
\title{Effects of Beyond Standard Model Physics on GRB Neutrinos}
\author{Reetanjali Moharana}
\email{reetanjali@phy.iitb.ac.in}
\affiliation{Department of Physics, Indian Institute of Technology Bombay, Mumbai - 400076, India}
\author{Debasish Borah}
\email{dborah@tezu.ernet.in}
\affiliation{Department of Physics, Tezpur University, Tezpur - 784028, India}

\begin{abstract}
The nondetection of neutrinos coming from Gamma Ray Bursts (GRBs) by the IceCube experiment has raised serious questions on our understanding of GRB's and the mechanism of neutrino flux production in them. Motivated by this and the need for a precise calculation for GRB neutrino flux, here we study the effects of beyond standard model physics on the GRB neutrino flux. In the internal shock model of GRB, high energy neutrinos are expected from muon, pion and kaon decays. Using the latest best fit neutrino oscillation parameters, we compute the expected flux on earth for standard as well as non-standard oscillation scenarios. Among the non-standard scenarios, we consider neutrino decay, pseudo-dirac nature of neutrinos and presence of one eV scale light sterile neutrino. Incorporating other experimental bounds on these new physics scenarios, we show that neutrino decay scenario can significantly alter the neutrino flux on earth from the expected ones whereas the corresponding changes for pseudo-dirac and sterile 
neutrino cases 
are 
moderate.
\end{abstract}

\pacs{12.60.-i, 98.70.Sa}
\maketitle

\section{Introduction}
Recent neutrino oscillation experiments have provided significant amount of evidence which confirms the existence of the non-zero yet tiny neutrino masses \cite{Fukuda:2001nk,Ahmad:2002jz,Ahmad:2002ka,Bahcall:2004mz,Nakamura:2010zzi}. The smallness of neutrino masses compared to electroweak scale can naturally be explained by seesaw mechanism \cite{Minkowski:1977sc,GellMann:1980vs,Yanagida:1979as,Mohapatra:1979ia,Schechter:1980gr}, the simplest version of which corresponds to the inclusion of three singlet right handed neutrinos into the standard model. Although these seesaw models can naturally explain the smallness of neutrino mass compared to the electroweak scale, we still do not have a complete understanding of the origin of neutrino mass hierarchies as suggested by experiments. Recent neutrino oscillation experiments T2K \cite{Abe:2011sj}, Double ChooZ \cite{Abe:2011fz}, Daya-Bay \cite{An:2012eh} and RENO \cite{Ahn:2012nd} have not only made the earlier predictions for neutrino parameters more precise, 
but also predicted non-zero value of the reactor mixing angle $\theta_{13}$. The latest global fit value for $3\sigma$ range of neutrino oscillation parameters \cite{GonzalezGarcia:2012sz} are as follows:
$$ \Delta m_{21}^2=(7.00-8.09) \times 10^{-5} \; \text{eV}^2$$
$$ \Delta m_{31}^2 \;(\text{NH}) =(2.27-2.69)\times 10^{-3} \; \text{eV}^2 $$
$$ \Delta m_{23}^2 \;(\text{IH}) =(2.24-2.65)\times 10^{-3} \; \text{eV}^2 $$
$$ \text{sin}^2\theta_{12}=0.27-0.34 $$
$$ \text{sin}^2\theta_{23}=0.34-0.67 $$ 
\begin{equation}
\text{sin}^2\theta_{13}=0.016-0.030
\label{globalfit}
\end{equation}
where NH and IH refers to normal and inverted hierarchy respectively.

Although the next generation neutrino oscillation experiments are expected to shed light on the origin of mass hierarchies as well as the Dirac CP phase, it is worth exploring if there exists an alternate experimentally verifiable way to understand some of the yet unresolved issues in neutrino physics. It will be even more exciting if such alternate ways can also confirm or rule out some of the well motivated beyond standard model frameworks which may or may not be seen in collider experiments. It turns out that the neutrino telescopes which have been designed to observe high energy cosmic rays, can be a promising setup to search for new physics.

The consequences of many such well motivated new physics scenarios on the observations of neutrino flux observed by neutrino telescopes have been studied by several groups \cite{Beacom:2002vi,Beacom:2003nh,Bhattacharya:2009tx,Bhattacharya:2010xj,Keranen:2003xd,Beacom:2003eu,Hooper:2004xr,Hooper:2005jp,Meloni:2006gv,Xing:2008fg,Esmaili:2009dz,Cowsik:2012qm,Meloni:2012nk,Baerwald:2012kc,Esmaili:2012ac,Pakvasa:2012db}. Motivated by these, here we pursue a similar study on the possibility of observing new physics at neutrino telescopes. In particular, we focus on high energy neutrinos coming from Gamma Ray Bursts (GRBs) and present an analysis of how the expected total flux of neutrinos at neutrino telescopes can change significantly by the presence of new physics. Among new physics scenarios, we consider neutrino decay, presence of one light sterile neutrino and pseudo dirac nature of neutrinos. 

Inside a GRB, shock accelerated protons may interact with low energy photons leading to the production of high energy mesons
 $p\gamma\rightarrow{\pi^{+,0}}X$ and these pions subsequently decay to high energy neutrinos ${\pi^{+}} \, \rightarrow \,
 {\mu^{+}}\nu_{\mu}, \, {\mu^{+}} \, \rightarrow \, {e^{+}}
 \bar\nu_{\mu}\nu_e$. This flux has already calculated in \cite{Vietri:1995hs,PhysRevLett.75.386,AlvarezMuniz:2000es,Gupta:2002zd,Guetta:2003wi,Bhattacharjee:2005nh,Gupta:2006jm,Murase:2005hy,Becker:2006gi,Waxman:2007kp,Murase:2008mr}. Recently IceCube collaboration has claimed to reach the sensitivity of detecting neutrino flux from GRBs at TeV energy \cite{Abbasi:2011qc}. The combined operation of Icecube 40 and 59 string for the time period of April 5, 2008 to May 2010 for GRB neutrinos has placed a tighter upper bound, 3.7 times below the theoretical predictions \cite{Abbasi:2012zw}. \cite{Hummer:2011ms,Li:2011ah,He:2012tq} have recalculated the neutrino flux from the 215 GRBs used by IceCube during their period of detection, and concluded that the neutrino flux predicted theoretically in the papers by IceCube collaborations is an overestimation. 

  However $p\gamma$ in GRBs can also produce high energy neutrons, and they will decay as $n\,\rightarrow p +e^-+\overline{\nu}_e$ to antineutrinos \cite{Moharana:2010su}. 
The other secondary products in $p\gamma$ interactions are 
$p\gamma \rightarrow{K^{+,0}}X$ where X can be either $\Lambda^0$, $\Sigma^0$ or $\Sigma^+$. 
Kaons decay to lighter mesons, leptons and neutrinos \cite{Baerwald:2010fk,Baerwald:2011ee,Moharana:2011hh}. In \cite{Baerwald:2010fk,Baerwald:2011ee} the dominant decay channel kaon to neutrino, $K^+ \rightarrow \mu^+\nu_{\mu}(63\%)$ was taken while we have considered all the channels of ${K^{+,0}}$ decaying to neutrinos \cite{Nakamura:2010zzi}. So one can find the total neutrino flux from GRBs has a contribution from these processes too.

Using the above mentioned possible origin of high energy neutrino flux from GRBs and the best fit values of neutrino oscillation parameters (\ref{globalfit}), we show that the individual neutrino flux can change significantly from the ones expected from standard oscillation paradigm. More specifically, the scenario of neutrino decay can change the expected flux to a great extent, wheareas the changes in the scenario of sterile neutrino and pseudo-dirac neutrino are somewhat moderate and should be detectable in future neutrino telescopes.

This paper is organized as follows: In section \ref{sec2}, we discuss the possible ways high energy neutrinos can originate in GRBs. Then we discuss the change in neutrino flux due to standard as well as non-standard oscillations in section \ref{sec3} and finally conclude in section \ref{sec4}.

\section{high energy neutrinos from gamma ray bursts}
\label{sec2}
In calculating the high energy neutrino flux from individual GRBs we have used the method as in ref. \cite{Moharana:2011hh}.
Frames of references are assigned as ``c'' for comoving or wind rest frame, ``p'' for proton rest frame. Quantities measured in the source rest frame are written without any subscript.
Shock accelerated high energy protons interacting with low enrgy photons will produce high energetic muons, pions, neutrons and kaons. We have used the low energy photon flux typically observed by $Swift$ in the energy range of 1 KeV to 10 MeV to calculate the neutrino flux from individual GRBs with break at
 $\epsilon_{\gamma}^b$ in the source rest frame related to the break energy
in the comoving frame
 $\epsilon_{\gamma,c}^b$ as $\epsilon_{\gamma}^b=\Gamma\epsilon_{\gamma,c}^b$.
\begin{equation}
\frac{dn_{\gamma}}{d\epsilon_{\gamma,c}}
=A \left\{ \begin{array}{l@{\quad \quad}l}
{\epsilon_{\gamma,c}}^{-\gamma_1} &
\epsilon_{\gamma,c}<\epsilon_{\gamma,c}^b\\{\epsilon_{\gamma,c}^b}^{\gamma_2-\gamma_1}
{\epsilon_{\gamma,c}}^{-\gamma_2} & \epsilon_{\gamma,c}>\epsilon_{\gamma,c}^b
\end{array}\right.
\label{photon}
\end{equation}
$\gamma_1 < 2$, and $\gamma_2 > 2$. The normalization constant $A$
is related to the internal energy density $U$ by,
\[
A=\frac{U{\epsilon_{\gamma,c}^b}^{\gamma_1-2}}{[\frac{1}{\gamma_2-2}-\frac{1}
{\gamma_1-2}]}
\label{A_g}
\]
The maximum energy of the shock accelerated protons in the GRB fireball 
can be calculated by comparing the minimum of the $p\gamma$ interaction time scale ($t_{p\gamma}$), p-synchrotron cooling time scale ($t_{syn}$) and dynamical 
time scale ($t_{dyn}$) of a GRB with the acceleration time scale ($t_{acc}$) 
of the protons as discussed in \cite{Gupta:2009gk}.
\begin{equation}
t_{acc}=min(t_{p\gamma},\,t_{dyn},\, t_{syn})
\end{equation}

 We have considered production of pions in $p\gamma$ interactions with two particle final states through the decay of resonant particle $\Delta^+$.
At the delta resonance both $\pi^{0}$ and $\pi^{+}$ have been assumed to be produced with equal probabilities. $\pi^{+}$ gets on the average
$20\%$ of the proton's energy.
The charged pions decay to muons and neutrinos. Finally the muons decay to electrons and neutrinos, antineutrinos. Each pion decay followed by muon decay gives two neutrinos, one antineutrino and one positron.
If the final state leptons share the pion energy equally then each neutrino carries $5\%$ of the initial proton's energy.
The specific parameters of GRBs can be denoted as the following way,
the fireball Lorentz factor $\Gamma_{300}= \Gamma/{300}$, photon luminosity $L_{\gamma,51}=L_{\gamma}/(10^{51} ergs \, /sec)$,
variability time $t_{v,-3}=(t_v/10^{-3}sec)$ are the important parameters of a GRB. The internal energy density $U$ relates to photon luminosity, $L_\gamma=4\pi {r_d}^2 \Gamma^2 c U$. $r_d=\Gamma^2ct_v$ is the internal shock radius. 

The total energy to be emitted by neutrinos of energy $\epsilon_{\nu,\pi}$ from photo-pion decay (considering muon and pion decay neutrinos together) 
in the source rest frame of a GRB is \cite{Gupta:2006jm},
\begin{equation}
 \epsilon_{\nu,\pi}^2\frac{dN_{\nu}}{d\epsilon_{\nu,\pi}}\approx\frac{3f_{\pi}}
{8\kappa}\frac{(1-\epsilon_e-\epsilon_B)}{\epsilon_e}E_{\gamma}^{iso}
 \left\{\begin{array}{l} 1 \hspace{2.cm} \epsilon_{\nu,\pi}<\epsilon_{\nu,\mu}^{s}\\
 \left(\frac{\epsilon_{\nu,\pi}}{\epsilon_{\nu,\mu}^{s}}\right)^{-2} \hspace{1.cm} \epsilon_{\nu,\pi}>\epsilon_{\nu,\mu}^{s}
\end{array}
 \right.
\label{totnupi}
\end{equation}
where $E_{\gamma}^{iso}$ is the total isotropic energy of the emitted
gamma-ray photons in the energy range of 1keV to 10MeV, which is
available from observations. It is the product of $L_{\gamma}$ with the duration of the prompt emission from the GRB. $\epsilon_e$ and $\epsilon_B$ are
 the energy fractions carried by electrons and the magnetic field respectively. 
The maximum energy of neutrinos from pion decay is approximately $5\%$ of 
the maximum energy of protons ($\epsilon_{p,max}$). $\epsilon_{\nu,\mu}^{s}$ presents the neutrino energy where muon synchrotron cooling starts. 

Ultrahigh energy neutrons are also produced in $p\gamma$ interactions along with pions and kaons. 
These neutrons (with Lorentz factor $\Gamma_n$) decay ($n\,\rightarrow p +e^-+\overline{\nu}_e$) to $\bar\nu_e$ with 
a decay mean free path $c\Gamma_n \bar{\tau}_n \,=\,10(\epsilon_n/EeV)$ Kpc.
 $\overline{\tau}_n\,=\,886$ seconds is the lifetime of a neutron in its rest frame and $\epsilon_n$ is its energy in the source rest frame. Assuming the probability of production of neutrons in resonant $p\gamma$  interactions to be half one would be able to calculate the fraction of a proton's energy lost to neutron production in the process $p\gamma\rightarrow \pi^+n$ at the $\Delta-$resonance \cite{Moharana:2010su} as,
\begin{equation}
f_n (\epsilon_p)= f_0^n \left\{\begin{array}{l@{\quad\quad}l}
 \frac{1.34^{\gamma_1-1}}{\gamma_1+1}\left(\frac{\epsilon_p}{\epsilon_{p,\Delta}^b}\right)^{\gamma_1-1}& {\epsilon_p>\epsilon_{p,\Delta}^b}\\
 \frac{1.34^{\gamma_2-1}}{\gamma_2+1}\left(\frac{\epsilon_p}{\epsilon_{p,\Delta}^b}\right)^{\gamma_2-1}&
{\epsilon_{p}< \epsilon_{p,\Delta}^b}
\end{array}
\label{fn}
\right.
\end{equation}
where $f_0^n=\xi_n\frac{4.5L_{\gamma,51}}{\Gamma_{300}^4 \,
t_{v,-3}{\epsilon_{\gamma,MeV}^b}}\frac{1}{\big
[\frac{1}{\gamma_2-2}-\frac{1}{\gamma_1-2}\big ]}$ and $\xi_n=0.8$. 
And the energy flux of antineutrinos of energy $\epsilon_{\bar{\nu},n}$  can be estimated with the neutron flux $({dN_n}/{d\epsilon_n})$ 
at the source rest frame travelling a distance $D_s$ \cite{Anchordoqui:2003vc} as,
\begin{equation}
\epsilon_{\bar{\nu},n}^2\frac{dN_{\bar{\nu}}}{d\epsilon_{\bar{\nu},n}}(\epsilon_{\bar{\nu},n})
=\left[\int\limits_{\frac{m_n\,\epsilon_{\bar{\nu},n}}{2\,\epsilon_0}}^{\epsilon_{n,max}}
\frac{d\epsilon_n}{\epsilon_n}\,\frac{dN_n}{d\epsilon_n}\left(1-e^{-\frac{D_sm_n}{\epsilon_n\,\overline{\tau}_n}}\right)\frac{m_n}{2\,\epsilon_0} \,\right] \times \epsilon_{\bar{\nu},n}^2.
\label{nu}
\end{equation}
$\epsilon_0$ is the mean energy of an antineutrino in the neutron rest frame.

 In $p\gamma$ interactions charged kaons ($K^+$) are produced through the following interactions $p\gamma \rightarrow K^+ \Lambda^0$ and $p\gamma \rightarrow K^+ \Sigma^0$. 
Although the cross-sections of these interactions are lower compared to photo-pion production \cite{Glander:2003jw}, at very high energy the total neutrino flux produced through 
kaon decay becomes higher than that from photo-pion decay \cite{Asano:2006zzb,Baerwald:2010fk}. The fractional energy transferred from shock accelerated protons to kaons has been calculated for all the channels of $K^+$ from both resonant and multiparticle production from p$\gamma$ interaction as \cite{Moharana:2011hh}. $K^+$ decay to secondary neutrinos by the following channels, 
$K^+ \rightarrow \mu^+\nu_{\mu}(63\%)$, $\pi^+\pi^0(21\%)$, 
$\pi^+\pi^+\pi^-(6\%)$, $\pi^0 e^+\nu_e(5\%)$, $\pi^0 \mu^+\nu_\mu(3\%)$ and 
$\pi^+\pi^0\pi^0(2\%)$.   
Due to their heavier mass $K^+$ cools at higher energy compared to muon and pion. The synchrotron cooling break energy in the kaon spectrum is at
\begin{equation}
 \epsilon_K^s= 2.2\times10^{9}\epsilon_e^{1/2}\epsilon_B^{-1/2}
L_{\gamma,51}^{-1/2}\Gamma_{300}^4t_{v,-3} \, {\rm GeV}.
\end{equation}
It is derived by comparing the decay and cooling time scales of kaons.
 Although the cross-section of kaon production is much less than that of 
pion production, the neutrino flux from kaon channel exceeds the flux from pion channel at very high energy due to the slower rate of cooling of kaons.

 Neutral kaons are produced in $p\gamma$ interactions with a cross-section $\sigma_{K^0,\Sigma} \approx 0.6 \times 10^{-30}$ $\text{cm}^2$ \cite{Lawall:2005np} 
 at the peak energy ${\epsilon_0}^{K^0}=1.45$ GeV and width $\delta\epsilon_{K^0}$ = 0.7 GeV. Half of the neutral kaons are assumed to be long lived kaons ($K_L^0$). $K^0$ can be produced in $p\gamma$ interactions with multiparticle final states ($p\gamma \rightarrow K^0_S \Lambda^0 \pi^+$, $ K^0_L \Lambda^0 \pi^+$ and $K^0_S \Sigma^+ \pi^0$). The cross-sections of these interactions are measured as $0.5 \times 10^{-30}\text{cm}^2$, $0.5 \times 10^{-30}\text{cm}^2$ and $0.2 \times 10^{-30}\text{cm}^2$ respectively \cite{Lawall:2005np}.
 $K_L^0$ decays through the following channels $\pi^+e^-\bar{\nu_e}(39\%)$, $\pi^+\mu^-\bar{\nu_\mu}(27\%)$, $\pi^0 \pi^0 \pi^0(21\%)$, and $\pi^+ \pi^- \pi^0(13\%)$. $K_S^0$ decays to two charged pions through this channel $\pi^+\pi^-(69\%)$. The pions finally decay to neutrinos and antineutrinos. The total neutrino flux from $K_L^0$ and $K_S^0$ can be calculated in the same way as $K^+$.\\

 The observed total neutrino flux with energy $\epsilon_{\nu}^{ob}$ on earth is,
\begin{equation}
\frac{dN_{\nu}^{ob}(\epsilon_{\nu}^{ob})}{d\epsilon_{\nu}^{ob}}=\frac{dN_{\nu}(\epsilon_{\nu})}{d\epsilon_{\nu}}\frac{1+z}{4\pi
D_s^2} 
\end{equation}
where $z$ is the redshift of the GRB.

 Our calculations are based on the standard internal shock model of GRBs. In internal shocks the shock radius $r_d$ is related to the bulk Lorentz factor $\Gamma$ and variability time $t_v$, $r_d=\Gamma^2 c t_v$, where $c$ is the speed of light. We have not assumed any relation among the GRB parameters $\Gamma$ and isotropic energy \cite{Liang:2009zi,Lu:2012zzc,Ghirlanda:2012gk} or peak luminosity and observed break energy in the low energy photon spectrum \cite{Ghirlanda:2012gk}. We have taken the following 4 set of GRB parameters to calculate the neutrino flux.
\begin{enumerate}
\item $\gamma_1=1$, $\gamma_2=2.2$, $L_\gamma=10^{53}$ erg/sec, $\Gamma=600$, $t_v=20$ msec, $\epsilon_\gamma^b=0.5 MeV$, $\epsilon_B$/ $\epsilon_e=1$, $f_{\pi}^0=0.09$ and $r_d=2.16 \times 10^{14}$ cm.\\
\item $\gamma_1=1.2$, $\gamma_2=2.5$, $L_\gamma=10^{53}$ erg/sec, $\Gamma=600$, $t_v=20$ msec, $\epsilon_\gamma^b=0.5 MeV$, $\epsilon_B$/ $\epsilon_e=10 ,
 (\epsilon_B=0.6, \epsilon_e=0.06)$, $f_{\pi}^0=0.17$ and $r_d=2.16 \times 10^{14}$ cm.\\ 
\item $\gamma_1=1.8$, $\gamma_2=2.01$, $L_\gamma=5\times10^{51}$ erg/sec, $\Gamma=130$, $t_v=25$ msec, $\epsilon_\gamma^b=0.5 MeV$, $\epsilon_B$/ $\epsilon_e=1$, $f_{\pi}^0=0.16$ and $r_d=1.26 \times 10^{13}$ cm.\\
\item $\gamma_1=1.2$, $\gamma_2=2.2$, $L_\gamma=10^{54}$ erg/sec, $\Gamma=1000$, $t_v=20$ msec, $\epsilon_\gamma^b=0.5 MeV$, $\epsilon_B$/ $\epsilon_e=1$, $f_{\pi}^0=0.12$ and $r_d=6 \times 10^{14}$ cm.
\end{enumerate}

\begin{figure}
\centering
\includegraphics{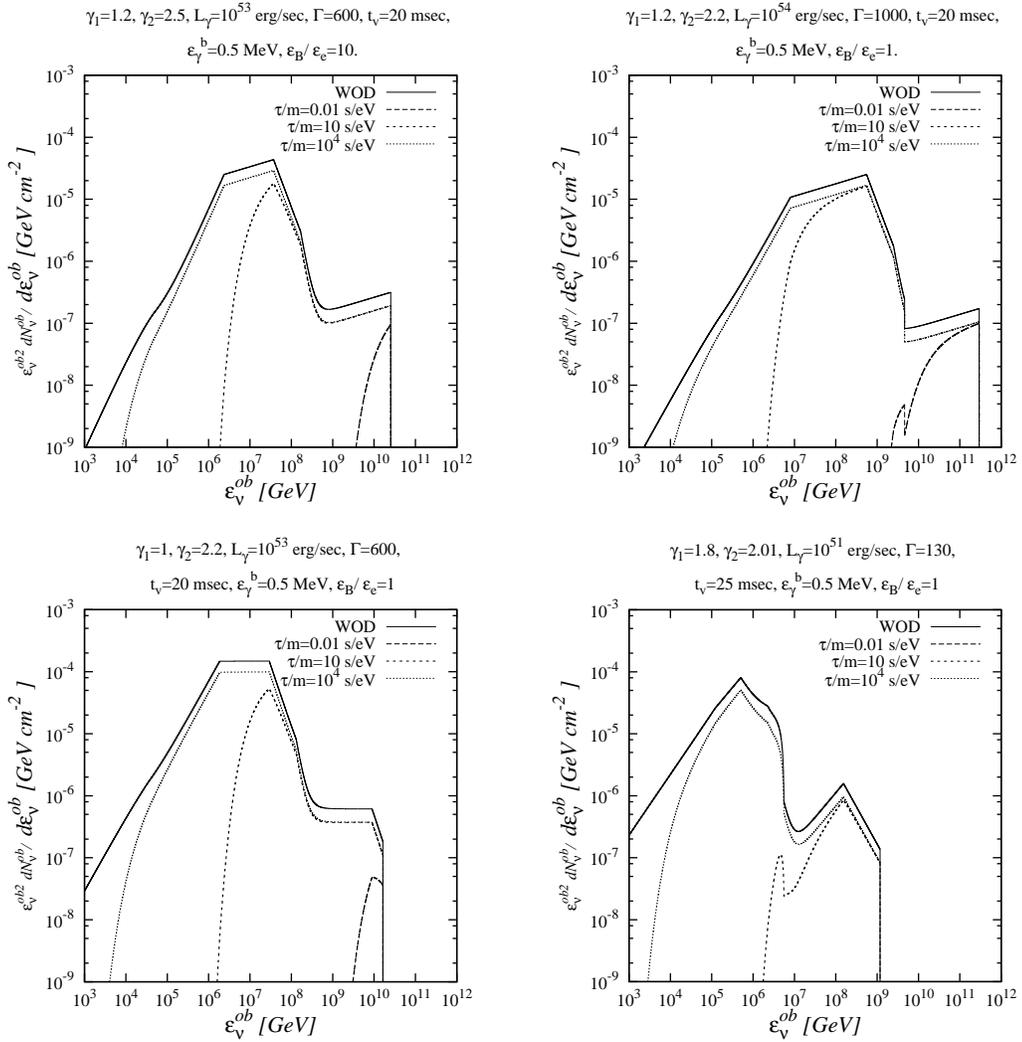} 
\caption{Total muon neutrino flux on earth for standard oscillation i.e., without decay (labelled as WOD) and for neutrino decay scenario in NH regime}   
\label{fig1}  
\end{figure}

\begin{figure}
\centering
\includegraphics{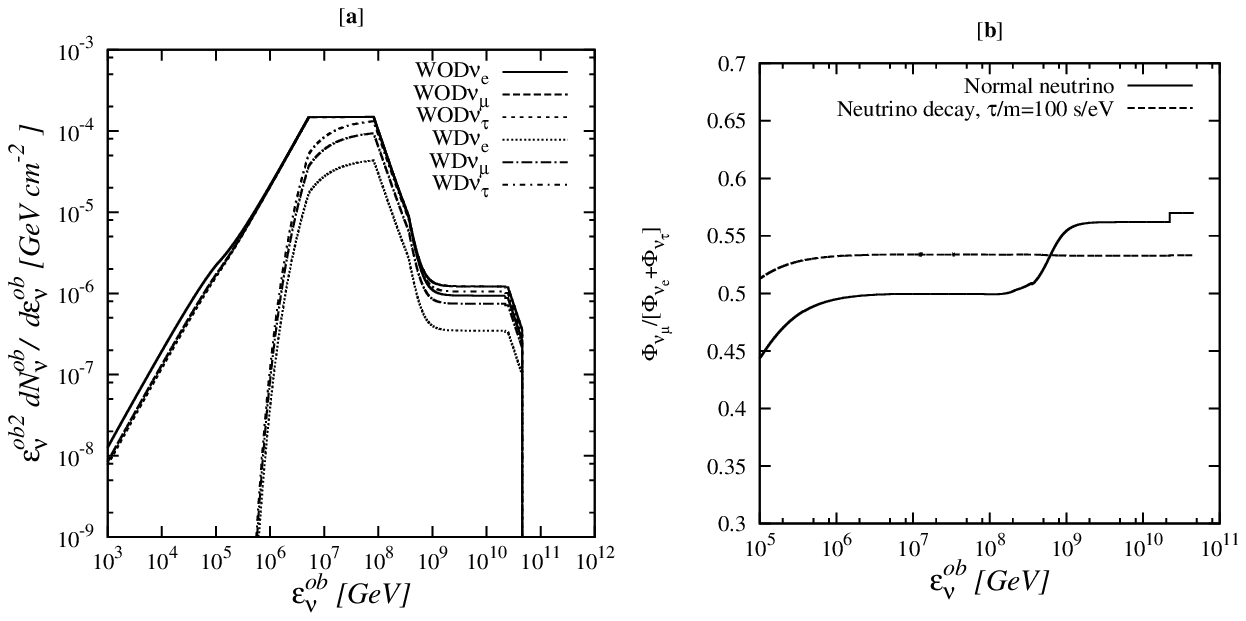}
\caption{$(a)$ Electron, muon and tau neutrino flavor composition for standard oscillation (WOD) and for neutrino decay scenario (WD) for $\tau_3/m_3 = \tau_2/m_2 = \tau_1/m_1 = 100$ for GRB 1. $(b)$ Neutrino flavor flux ratio for the GRB 1}
\label{fig2}
\end{figure}
\section{Effect of new physics on GRB neutrinos}
\label{sec3}

\subsection{Standard Neutrino Oscillation}
Neutrino oscillation data clearly indicate the smallness of three Standard Model
neutrino masses \cite{Fukuda:2001nk, Ahmad:2002jz, Ahmad:2002ka, Bahcall:2004mz} which can be naturally explained 
via see-saw mechanism \cite{Minkowski:1977sc, GellMann:1980vs, Yanagida:1979as, Mohapatra:1979ia}. Without using any particular type of seesaw, here we use the most general neutrino mixing MNS matrix \cite{Maki:1962mu} and the standard vacuum oscillation probability given by 
\begin{equation}
P(\nu_{\alpha} \rightarrow \nu_{\beta} ; L) = \delta_{\alpha \beta} - \sum_{j \neq k}U^*_{\alpha j} U_{\beta j} U_{\alpha k} U^*_{\beta k} (1 - e^{-i \Delta E_{jk} L}) 
\end{equation}
where $U_{\alpha i}$ is an element of the the MNS matrix, $\alpha, i$ denoting flavor and mass eigenstates respectively. For cosmological distances like the typical distance of a GRB from earth we can assume the limit $L \rightarrow \infty $ which simplifies the above expression for probability to 
\begin{equation}
P (\nu_{\alpha} \rightarrow \nu_{\beta} ) = \sum_j \lvert U_{\alpha j} \rvert^2 \lvert U_{\beta j} \rvert^2
\end{equation}
The role of standard neutrino oscillation on ultra high energy neutrino flux from GRB's were studied earlier in \cite{Learned:1994wg,Athar:2000yw}. These studies concluded that the neutrino flavor ratio $1:2:0$ at source would reduce to $1:1:1$ at earth due to oscillations. The neutrinos while coming from a distance of z will undergo oscillation. We have taken the standard oscillation parameters along with the recent calculated value of $\text{sin}^2{2\theta_{13}}=0.1$, where $\theta_{13}$ is the neutrino mixing angle 
measured by Double ChooZ \cite{Abe:2011fz}, Daya-Bay \cite{An:2012eh} and RENO \cite{Ahn:2012nd} collaborations and other oscillation parameters from global fit data \cite{GonzalezGarcia:2012sz}.

\subsection{Neutrino Decay}
If the neutrino mass eigenstates are hierarchical, then a higher mass eigenstate can decay into a lower mass eigenstate. The role of such neutrino decay on neutrino flavor flux was studied in \cite{Beacom:2002vi}. Here we consider the simplest possible situation where the heavier mass eigenstate completely decays into the lightest mass eigenstate which is kinematically stable. Thus in case of normal hierarchy (NH)$ m_{\nu_3} > m_{\nu_2} >m_{\nu_1}$, the mass eigenstate ratio at earth will be $1:0:0$. Hence the flavor ratio at earth will be $0.67:0.26:0.07$ for NH .

For incomplete decay, we have to include the decay factor in the expression for probability of oscillation. This decay factor which accounts for the depletion in neutrino flux due to the decay of mass eigenstate $m_i$ with rest-frame lifetime $\tau_i$ and energy $E$ propagating over a distance $L$, is $\exp(-\frac{L}{E}\frac{m_i}{\tau_i})$. The expression for neutrino flux at earth in this case becomes 
\begin{equation}
\phi_{\nu_{\alpha}}(E) = \sum_i \sum_{\beta} \phi^0_{\beta}(E)\lvert U_{\beta i} \rvert^2 \lvert U_{\alpha i} \rvert^2 e^{-\frac{L}{E}\frac{m_i}{\tau_i}}
\end{equation}
where $\phi^0_{\beta}$ denotes the flux of neutrino flavor $\beta$ at source. Taking the flavor ratio at source to be $x:y:z$ and using the oscillation parameters as in \cite{GonzalezGarcia:2012sz}, we compute the flux for different neutrino flavors at earth.
 For the purpose of our calculation we consider three cases $\tau_3/m_3 = \tau_2/m_2 = \tau_1/m_1 = 0.01, 10, 10^4 \; \text{s}/\text{eV} $ respectively as shown in figure \ref{fig1}. Also, the electron, muon and tau flavor composition of neutrino flux on earth is shown for one specific choice of GRB parameters as well as $\tau/m$ in figure \ref{fig2} $(a)$. We also show the flavor ratio of muon type to electrom plus tau type neutrinos in figure \ref{fig2} $(b)$ for the same choice of GRB parameters.

\begin{figure}
\centering
\includegraphics{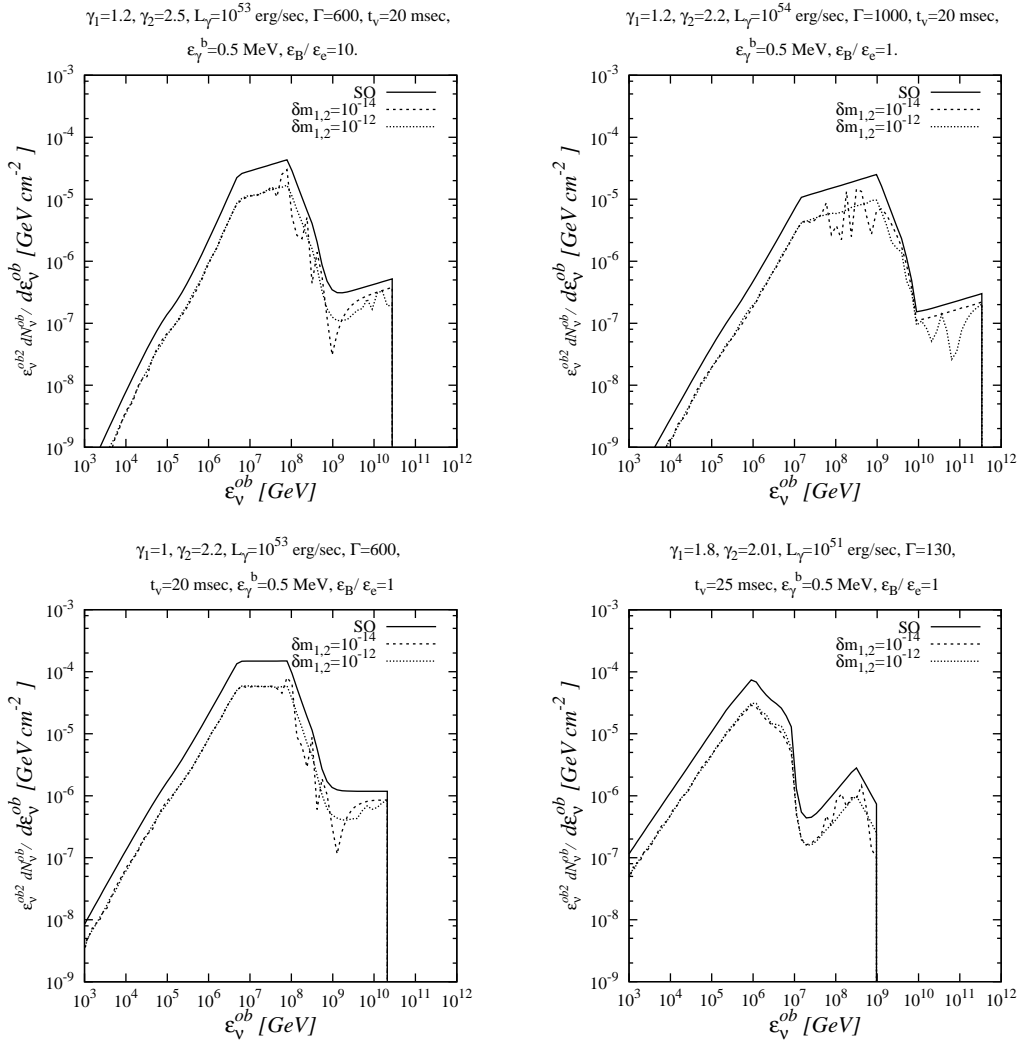} 
\caption{Total muon neutrino flux on earth for standard oscillation (SO) and for Pseudo dirac neutrino scenario}  
\label{fig3}  
\end{figure}

\begin{figure}
\centering
\includegraphics{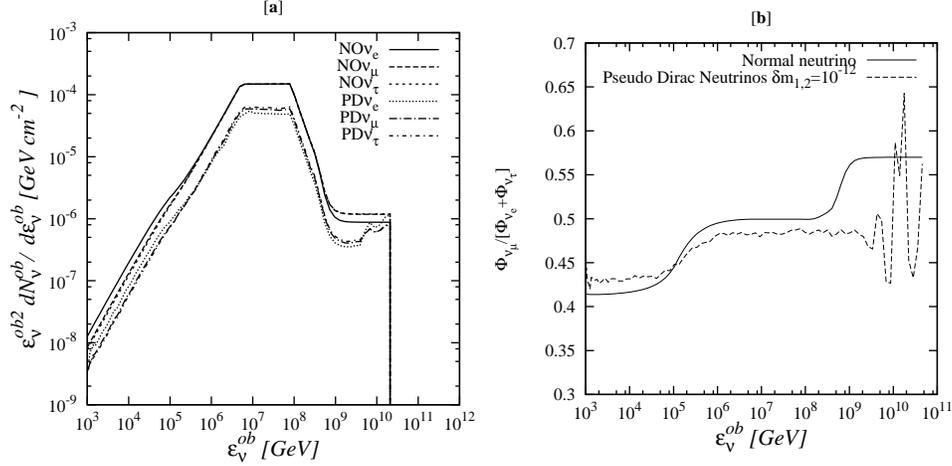}
\caption{$(a)$ Electron, muon and tau flavor composition of neutrino flux on earth for standard oscillation (SO) and for Pseudo dirac neutrino scenario (PD) $\delta m^2_{1,2} = 10^{-12}$ for GRB 1. $(b)$  Neutrino flavor flux ratio for the GRB 1}
\label{fig4}
\end{figure}
\begin{figure}
\centering
\includegraphics{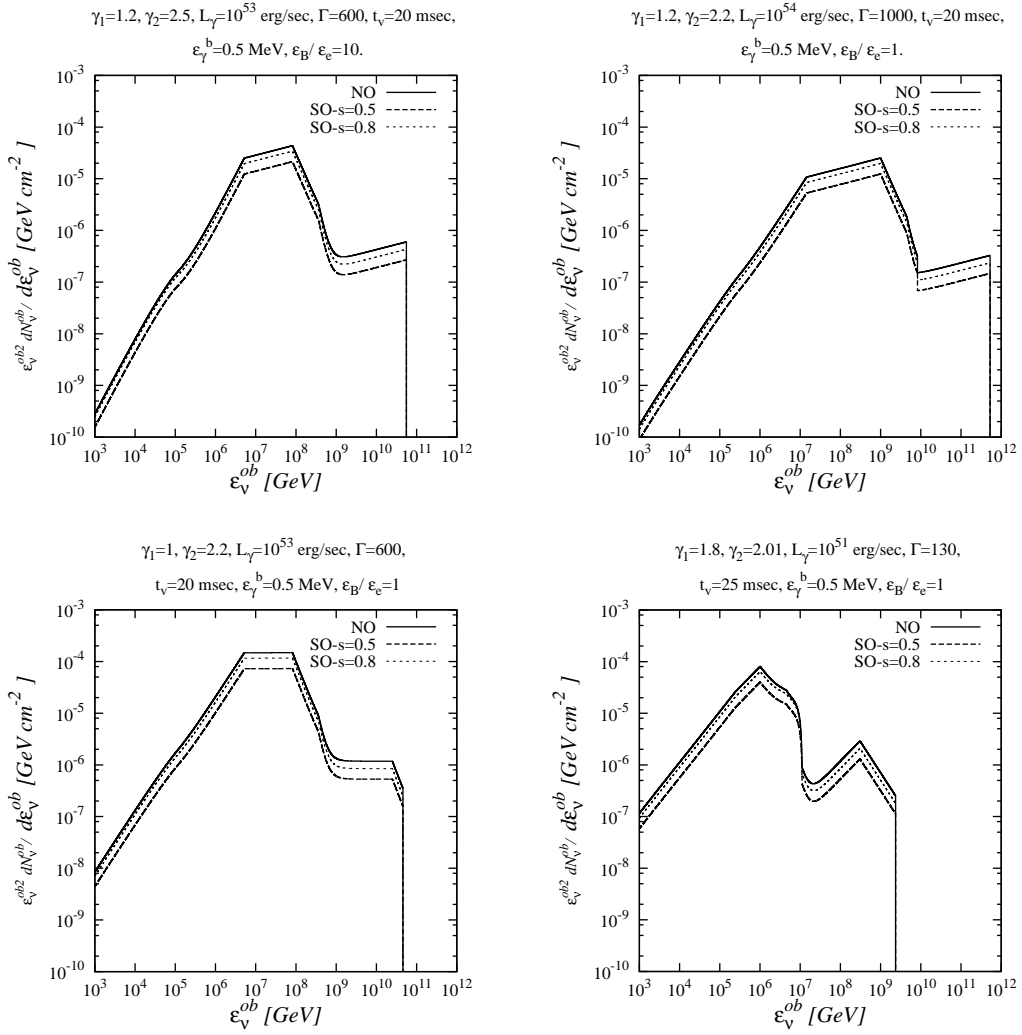} 
      \caption{Total muon neutrino flux on earth for normal oscillation (SO) and for one extra sterile neutrino (SN) case}
\label{fig5}  
\end{figure}

\begin{figure}
\centering
\includegraphics{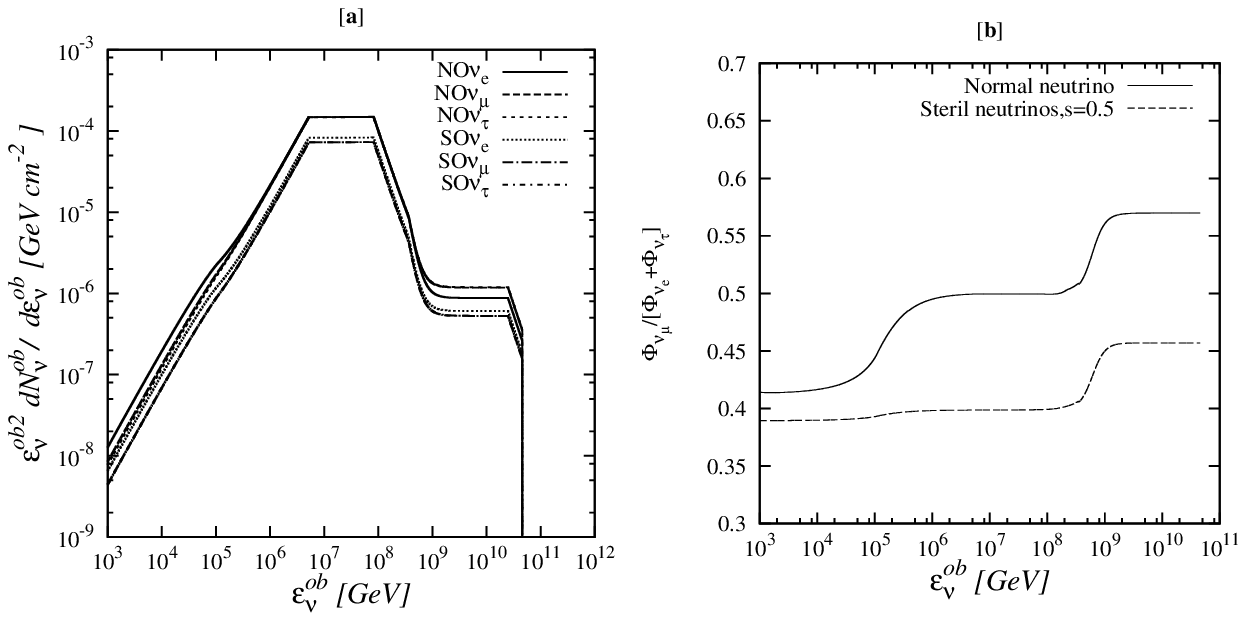}
\caption{$(a)$ Electron and muon flavor composition of neutrino flux on earth for normal oscillation (SO) and for one extra sterile neutrino (SN) case for $\it {S}_i=0.5$ for GRB 1. $(b)$ Neutrino flavor flux ratio for the GRB 1}
\label{fig6}
\end{figure}

\subsection{Pseudo Dirac Neutrinos}
Instead of a being purely Dirac with a mass $m_D$, neutrino can be a mixture of two almost degenerate Majorana neutrinos. Please see \cite{Beacom:2003eu} for the role of pseudo Dirac neutrinos in ultra high energy neutrino flux and references therein for earlier works on pseudo Dirac neutrinos. In such a case the majorana mass terms $m_L, m_R \ll m_D$ and the the mass splitting is $\delta m^2 \simeq 2m_D (m_L +m_R) $. Thus the neutrino mass and flavor basis become $(\nu^+_1, \nu^+_2, \nu^+_3, \nu^-_1, \nu^-_2, \nu^-_3)$ and $(\nu_e, \nu_{\mu}, \nu_{\tau}, \nu'_e, \nu'_{\mu}, \nu'_{\tau})$ respectively. As shown in \cite{Beacom:2003eu}, the neutrino flavor conversion probability can have a form as simple as 
\begin{equation}
P_{\alpha \beta} = \sum_{\alpha} \sum^3_{j=1} \lvert U_{\alpha j} \rvert^2 \lvert U_{\beta j} \rvert^2 \cos^2(\frac{\delta m^2_j L}{4E})
\end{equation}
The new constribution coming from pseudo Dirac nature of neutrinos will be negligible until $E/L$ becomes of the order of $\delta m^2_j$. $\delta m^2$ can be as large as $10^{-12} \text{eV}^2$ for $\nu_{1,2}$ and as large as $10^{-4} \text{eV}^2$ for $\nu_3$. Taking the initial neutrino flavor ratio at source to be $x:y:z$ and using the same neutrino oscillation parameters as in \cite{GonzalezGarcia:2012sz} we find the probability of detecting individual flavors on earth. Here we use $\delta m^2_{1,2} = 10^{-14}, 10^{-12}\; \text{eV}^2,\; \delta m^2_3 = 10^{-6} \;\text{eV}^2 $ for the purpose of our calculation. $L$ can be taken to be $100 \text{Mpc} $ which is the typical distance of a GRB from earth. The consequences of this pseudo-dirac nature of neutrinos on the muon neutrino flux on earth is shown in figure \ref{fig3}. The electron, muon and tau flavor composition of neutrino flux for one specific GRB parameter and $\delta m^2$ is shown in figure \ref{fig4} $(a)$. We also show the flavor ratio of muon type to electrom plus tau type neutrinos in figure \ref{fig4} $(b)$.

\subsection{Presence of Sterile Neutrinos}
Precision measurement of the Z boson decay width  restricts the number of standard model neutrinos to three. 
However we can still have an eV scale neutrino which has no coupling to the Z boson and hence called sterile. 
The presence of such a sterile neutrino is also compatible with the recent Wilkinson Microwave Anisotropy Probe (WMAP) 
collaboration data \cite{Komatsu:2010fb} with $95 \% $ allowed mass range $ < 0.48 \; \text{eV} $ \cite{Hamann:2010bk}.
 
 The effect of sterile neutrino on ultra high energy neutrino flux was studied earlier in \cite{Keranen:2003xd}.
 We follow their approach and use the recent oscillation data \cite{GonzalezGarcia:2012sz} to see the effect of such sterile neutrinos on 
 the neutrino flux coming from GRBs. As considered in \cite{Keranen:2003xd}, we also consider the presence of a sterile neutrino which 
 is almost degenerate with the lightest active neutrino mass eigenstate. Interestingly, if the mass difference is $\delta m^2 \le 10^{-11} \; \text{GeV} $ 
 then there is no experimental constraint on the mixing angles of this sterile neutrino with the three active ones. As outlined in \cite{Keranen:2003xd},
 the simplified oscillation probability for $3+1$ neutrino scheme is 
 \begin{equation}
P_{\alpha \beta} = S_1 \lvert U_{\alpha 1} \rvert^2 \lvert U_{\beta 1} \rvert^2 +S_2\lvert U_{\alpha 2} \rvert^2 \lvert U_{\beta 2} \rvert^2 + S_3\lvert U_{\alpha 3} \rvert^2 \lvert U_{\beta 3} \rvert^2
\end{equation}
where $S_i = \cos^4 \phi_i + \sin^4 \phi_i $ and $ \phi_i$ is the mixing angle between the active neutrino mass eigenstate
$\nu_i$ and the sterile state $\nu_s $. In general $S_i$ factors can vary in the range $\frac{1}{2} \le S_i \le 1$. 
Taking the flavor ratio at source to be $x:y:z$ and using the neutrino oscillation data from \cite{GonzalezGarcia:2012sz} 
we find the probability of individual neutrino flavors at earth.
For zero mixings that is, $\phi_i =0$, the flavor ratios correspond to the standard oscillation as discussed earlier. 
To show the extent to which the presence of sterile neutrino can affect the flavor fluxes we plot the total muon neutrino flux on earth for two different cases $S_i =0.5, 0.8$ and compare with the standard predictions as shown in figure \ref{fig5}. The maximum value of this parameter $S = 1$ reproduces the standard oscillation results as seen from figure \ref{fig5}. We also show the electron, muon and tau flavor composition of neutrino flux on earth in figure \ref{fig6} $(a)$. We also show the flavor ratio of muon type to electrom plus tau type neutrinos in figure \ref{fig6} $(b)$. It should be noted that a larger $( \sim \text{eV}^2 )$ mass difference
between the active and the sterile mass eigenstate has other interesting motivations like explaining the anomalies found in MiniBooNE 
\cite{AguilarArevalo:2010wv} and LSND \cite{Aguilar:2001ty} which we do not attempt to explain here. In case of larger mass difference, 
the mixing angles between active and sterile neutrinos will be tightly constrained and hence will not affect the flavor ratios of ultra high energy neutrinos.

\section{Results and Discussion}
\label{sec4}
We have presented an analysis of the effect of three different new physics scenarios namely, neutrino decay, pseudo Dirac nature and presence of one light sterile neutrino on the flux of high energy neutrinos coming from GRB's. We first calculate the GRB neutrino flux from various sources like pion, muon, neutron and kaon decays by choosing four sets of GRB parameters. We then incorporate the standard neutrino oscillation between all three flavors (electron, muon and tau) and calculate the total muon flux on earth as well as the flavor composition. We show that different choices of GRB parameters can give rise to different neutrino flux on earth.

After calculating the neutrino flux on earth using standard neutrino oscillation physics, we then incorporate the three beyond standard model frameworks mentioned above to calculate the total muon flux on earth as well as the flavor composition. We find that for neutrino decay scenario, the changes in neutrino flux from the standard oscillation case can be very significant and could give rise to an explanation of the present non-detection  of GRB neutrinos at IceCube experinet. However for other two beyond standard model physics scenarios we consider namely, for pseudo Dirac and sterile neutrino cases, the changes are moderate. 

For the standard oscillation case, we get three plateau regions in the flavor ratio plot as can be seen in part (b) of figure \ref{fig2}, \ref{fig4} and \ref{fig6}. At lower energy, the neutron decay channel dominates while the muon damping corresponds to the transition to the second plateau region. As can be seen from the same figures, for the three beyond standard model scenarios the flavor ratio plot changes from the standard oscillation case. 

\section{Acknowledgement}
We would like to thank Prof. Nayantara Gupta, RRI Bangalore and Prof. S. Umasankar, IIT Bombay for comments and suggestions.


\end{document}